\documentclass[english]{article}
\usepackage[T1]{fontenc}
\usepackage[latin1]{inputenc}
\usepackage{a4wide}
\usepackage{amsmath}
\usepackage{graphicx}
\usepackage{amssymb}

\makeatletter

\providecommand{\tabularnewline}{\\}

 \newcommand{\lyxaddress}[1]{
   \par {\raggedright #1 
   \vspace{1.4em}
   \noindent\par}
 }

\usepackage{epsf}
\usepackage{rotate}
\usepackage{epsfig} 
\usepackage{graphicx} 
\usepackage{graphics} 

\usepackage{babel}
\makeatother
\begin{document}

\title{Two hard spheres in a spherical pore: Exact analytic results in two
and three dimensions}

\author{Ignacio Urrutia}

\maketitle

\lyxaddress{\begin{center}Departamento de Física, Comisión Nacional de Energía
Atómica,\\ Av. Gral. Paz 1499 (RA-1650) San Martín, Buenos Aires,
Argentina\\ iurrutia@cnea.gov.ar\end{center}}

\begin{abstract}
The partition function and the one- and two-body distribution functions
are evaluated for two hard spheres with different sizes constrained
into a spherical pore. The equivalent problem for hard disks is addressed
too. We establish a relation valid for any dimension between these
partition functions, second virial coefficient for inhomogeneous systems
in a spherical pore, and third virial coefficients for polydisperse
hard spheres mixtures. Using the established relation we were able
to evaluate the cluster integral $b_{2}(V)$ related with the second
virial coefficient for the Hard Disc system into a circular pore.
Finally, we analyse the behaviour of the obtained expressions near
the maximum density.
\end{abstract}

\section{Introduction}

The Hard Spheres (HS) and Hard Disks (HD) systems have attracted the
interest of many physicists owing to they constitute prototypical
simple fluids. Hard-core models, such as HD and HS are paradigmatic
entropy driven systems. The search of their accurate equation of state
is a long-standing problem of great importance in statistical mechanics
\cite{key-15,eos_HD,eos_HS,Eisen}. The evaluation of virial coefficients
\cite{key-15,HR,Boltz,key-5,HS_g,key-9} and the analysis of phase
transitions \cite{Eisen,HD_homogtrans} are some of the principal
subjects. The monodisperse systems as well as bidisperse and polydisperse
ones have been extensively studied.

Recently special attention was paid to systems of few HS and HD confined
in small vessels. The study of these inhomogeneous systems has shed
light on aspects of loss of ergodicity \cite{Erg}, freezing and glass
transitions \cite{Nemeth}, thermodynamic second law \cite{2ndlaw},
and other fundamental questions of statistical mechanics and thermodynamic
\cite{HSmelt,HD_2inBoxMD}.

Exact and approximate analytic properties of hard spherical bodies
at different dimensions have been very important in the evolution
of Free Energy Density Functional Theories for HS \cite{Rosen,Taraz}.
Besides its own relevance the HS system is the starting point of several
approximate theories of liquids \cite{Rosen0,BookHansen}, therefore
exact analytical results become still more interesting. Even though
the apparent simplicity of HS and HD fluids, a few exact analytical
results are known at present. For homogeneous systems the third and
fourth virial coefficients for HS and HD have been calculated on the
monodisperse \cite{HS_g,HS_g0,HS_g1,B4_evenD,B4_oddD} and polydisperse
systems \cite{HD_vir3Poly,HS_vir3Poly,HS_vir4Poly}, numerical results
for higher order terms have also been obtained \cite{HS_virBid,HS_virBid2}.
In addition, for inhomogeneous systems of HS into a spherical cavity
the analytic expression for the second virial coefficient is known
\cite{HS_inhom}. Until present the exact canonical partition function of
few bodies in a pore has been solved for two HD \cite{HD_2inBox} and
recently for two HS \cite{HS_2inBox}, in a rectangular box. Such results
for equal sized particles were compared with dynamical simulation
approaches \cite{HD_2inBoxMD,HD_2inBoxMDbis,HDp_2inBoxMD}. The three HD
in a rectangular box problem was analyzed \cite{HD_3inBox} but the
complete integration of its configuration integral (CI) could not be
performed. Exact analytic results exist for the zero dimensional limit,
which corresponds to a so narrow cavity that is able to contain one
particle at most. Also, several exact results have been obtained for the
Hard Rod system (HR) the 1D version of HS \cite{HR,HRbis} which actually
may be considered completely solved.

The aim of the present work is to evaluate the canonical partition
function for two spherical particles in a spherical pore at two and
three dimensions. Expressions for the free energy and one particle
density are also derived. In Sec. \ref{sec:The-In-Out-Diagram} of this
work we show that in the canonical ensemble a spherical pore containing
HS or HD can be seen as another particle. We establish a relation between
the CI of the canonical partition function for \textit{N} polydisperse
hard spherical particles and the respective CI for \textit{N-1} particles
inside a spherical vessel. In Sec. \ref{sec:Two-bodies-in} we evaluate
the canonical partition and density distribution functions for two
HD and HS into a spherical pore. The obtained expressions apply to
the non additive system of unequal sized particles in a pore, which
includes the more usual and restrictive additive system. The analysis
of the CI expressions and the thermodynamic properties of the system
are shown in Sec. \ref{sec:Results}. Finally in Sec. \ref{sec:Conclusions}
we present our conclusions.

\section{The In-Out Diagram relation\label{sec:The-In-Out-Diagram}}

In this section we will refer to non additive HS, but the discussion
also applies to HD and to the equivalent system in any dimension.
We are interested in a system with fixed number of particles (\textit{N})
and temperature, then we will consider the canonical ensemble. In such
ensemble the partition function factorizes into CI and a trivial kinetic
terms. The CI of the system is characterized by the set of hard repulsion
distances between each pair of particles $\{ d_{ij}\}$. The system is
additive only if the repulsion distances are $d_{ij}=R_{i}+R_{j}$ for 
positive radii of particles $\{ R_{i}\}$. When the above relation is not
fullfilled or any other particular assumption is made on the set
$\{ d_{ij}\}$ the system is non additive. We will analyze the CI for a
\textit{N}-HS system in an open (and virtually infinite) volume ($Q_{N}$).
In order to obtain a non null contribution to the integral, each pair of
particles $ij$ must fulfil $r_{ij}>d_{ij}$, where $r_{ij}$ is the
distance between particles $ij$. The above relation between $r_{ij}$ and
$d_{ij}$ is a consequence of the exponential Boltzmann's factor.
Then we have
\begin{equation}
Q_{N}=\int\cdots\int\,\prod_{<ij>}e_{ij}\, dr_{1}\ldots dr_{N}\;,
\label{eq:QN}\end{equation}
\begin{equation}
\begin{array}{ccc}
e_{ij} & =exp\left(-\phi_{ij}(r_{ij})\right)= & \left\{ \begin{array}{cl}
0\; & \mathrm{if}\:\, r_{ij}\leq d_{ij}\;\mathsf{(overlap)}\;\\
1\; & \mathrm{if}\:\, r_{ij}>d_{ij}\;\mathsf{(no\: overlap)}\;,\end{array}\right.\end{array}
\label{eq:ebond}\end{equation}
 where $\phi_{ij}(r_{ij})$ is the hard core spherical potential for
the pair of particles $ij$. In virial series and related subjects
this factor is usually named $\widetilde{f}$, however in the framework
of canonical partition function it appears naturally as an exponential
function. Then we prefer to name it $e$ or $e_{ij}$ function or
simply an $e$-bond, which will be plotted as a dashed segment. Such
a $Q_{N}$ can be expressed as a Ree-Hoover's graph (also called modified
star graph) \cite{HS_g,HS_g0}
\begin{equation}
Q_{N}=\raisebox{-20pt}{\psfig{figure=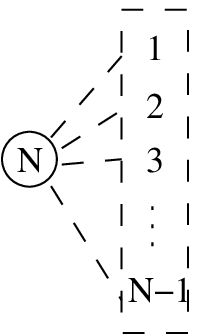,width=1.3cm}}\;,
\label{eq:QNgraph}\end{equation}
 where the \textit{N}-th particle at Eq. (\ref{eq:QNgraph}) was drawn
separately and its $e$-bond appears explicitly. The remaining (\textit{N-1})-particle
system with $e$-bonds between pairs is outlined as a rectangular
box. The relation between the Mayer $f$ function (or $f$-bond) and
$e$ is given by
\begin{equation}
e_{ij}=1+f_{ij}\;,
\label{eq:ef}\end{equation}
\begin{equation}
\begin{array}{ccc}
f_{ij} & = & \left\{ \begin{array}{ll}
-1\; & \mathrm{if}\:\, r_{ij}\leq d_{ij}\;\mathsf{(overlap)}\;\\
0\; & \mathrm{if}\:\, r_{ij}>d_{ij}\;\mathsf{(no\: overlap)}\;.\end{array}\right.\end{array}
\label{eq:fbond}\end{equation}
We can transform each $e$-bond that links the pairs $Ni$ ($i\,:1,...,N-1$)
through the usual bond relation Eq. (\ref{eq:ef}) performing the
decomposition over particle $N$
\begin{equation}
Q_{N}=\raisebox{-20pt}{\psfig{figure=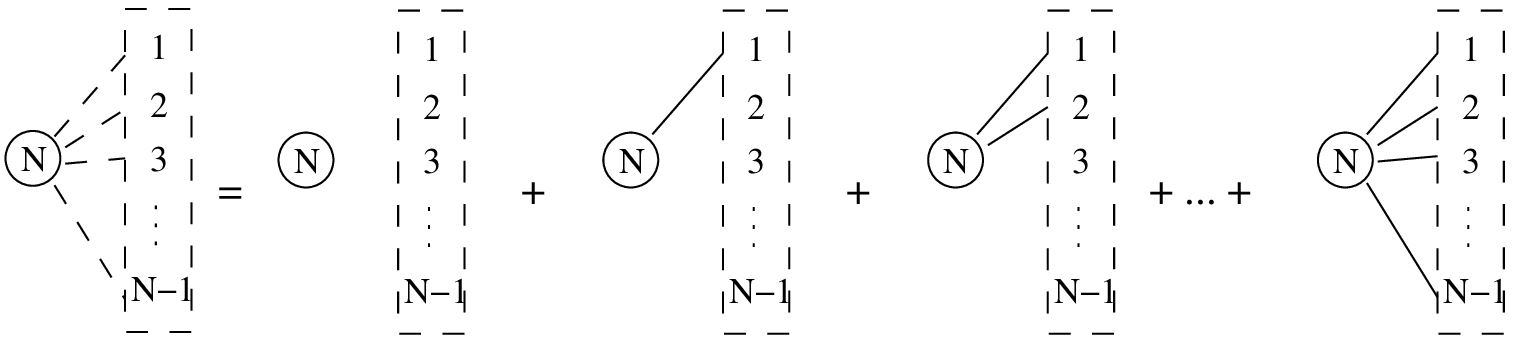,width=8.5cm}}\;.
\label{eq:QNdecomp}\end{equation}
In this representation all the equivalent graphs on the left of Eq.
(\ref{eq:QNdecomp}) that takes into account permutations of the \textit{N-1}
particles were omitted for clarity. It is possible to map this general
problem to the \textit{N}-HS additive system defined by the set of
particle radii $\{ R_{i}\}$ through the assignment $d_{ij}=R_{i}+R_{j}$.
We define the additive/subtractive generalization of the additive
system by the assignment $d_{ij}=\left|R_{i}\pm R_{j}\right|$, and
specifying which set of pairs assume the minus sign. If we are interested
in a system of additive spheres contained in a spherical vessel, as
is our case, the exclusion distances between the center of the pore
(which is the N-particle) and the center of each \textit{i}-particle
is $d_{iN}=(P-R_{i})$, where $P$ is the radio of the spherical pore.
We name to this specific additive/subtractive $\{ d_{ij} \} $ as an
additive-in-pore system.

Through the assignment of $\{ d_{ij}\}$ for an additive-in-pore system,
the last graph of Eq. (\ref{eq:QNdecomp}) may be interpreted as the
CI of (\textit{N-1}) additive HS in a pore ($Q_{(N-1)}^{P}$), due to the
condition $r_{iN}\leq d_{iN}$ with $1\leq i\leq N-1$ imposed by the
$f$-bonds in Eq. (\ref{eq:fbond}). Therefore $\{ d_{iN}\}$ in that graph
should be named inclusion (and not exclusion) distances. More precisely,
last graph is $V_{\infty}$ times $Q_{(N-1)}^{P}$, being $V_{\infty}$
the volume of the infinite space,
\begin{equation}
\begin{array}{ccl}
Q_{(N-1)}^{P} & = & V_{\infty}^{-1}\int\cdots\int\,\prod_{<iN>}f_{iN}\,\prod_{<ij>}e_{ij}\, dr_{1}\ldots dr_{N-1}dr_{N}\;,\\
\\ & = & \begin{array}{l}
\int_{P}\cdots\int\,\prod_{<ij>}e_{ij}\, dr_{1}\ldots dr_{N-1}\;,
\end{array}\end{array}
\label{eq:QN-1inPore}\end{equation}
where the integration domains depend on pore size $P$. From now on,
last graph of Eq. (\ref{eq:QNdecomp}) will be called the (\textit{N-1})-in-pore
graph, even in the more general framework of non additive systems.
If we have $N-1$ equal (and additive) hard spheres of radio $\sigma/2$
into a pore of radio $P$, on the additive-in-pore picture we assign
$d_{iN}=(P-\sigma/2)$. The same system can be analyzed as an entirely
additive system with $d_{iN}=(P-\sigma)+\sigma/2$, if $P>\sigma$.
We can obtain a Mayer type decomposition by transforming iteratively
each $e$-bond in Eqs. (\ref{eq:QNgraph}, \ref{eq:QN-1inPore}) by
an $f$-bond.

\section{Two bodies in a pore\label{sec:Two-bodies-in}}

In this section we will apply the in-out relation for three bodies.
In Eq. (\ref{eq:tri1}) we perform the simple decomposition over the
particle $P$ and show the relation between the \textit{2}-particles-in-pore
graph and the Mayer type graph
\begin{equation}
\raisebox{-20pt}{\psfig{figure=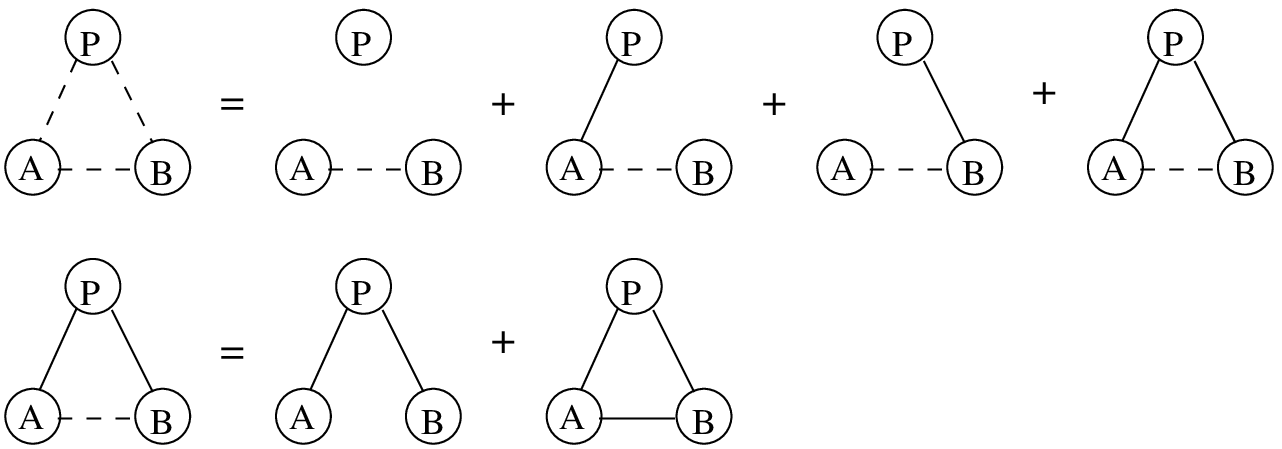,width=8.0cm}}\;.
\label{eq:tri1}\end{equation}
All the integrals in Eq. (\ref{eq:tri1}) may be directly evaluated, the
easiest are the simply connected graphs that become factorable
\cite{BookHill}. In the evaluation of different graphs we will use the
notation of \cite{HS_vir4Poly}. The exclusion/inclusion distances between
pair of spheres are $AB$, $PA$, $PB$. Being that $\{ AB, PA, PB \}$ are
completely independent we deal with a non additive system. Now we will
briefly analyze the two bodies or one body in a pore problem. Assuming that
$PA$ is the exclusion/inclusion distance between $A$ and $P$, we have for
HS
\begin{equation}
\raisebox{-1pt}{\psfig{figure=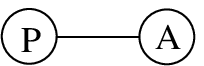,width=1.5cm}}=Sp(PA)\:,
\label{eq:2cont}\end{equation}
\begin{equation}
\raisebox{-1pt}{\psfig{figure=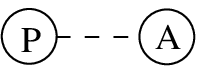,width=1.6cm}}=V_{\infty}-Sp(PA)\:,
\label{eq:2dash}\end{equation}
\begin{equation}
Sp(R)=\frac{4\pi}{3}R^{3}\:,
\label{eq:SpVol}\end{equation}
 where $Sp(R)$ is the volume of the sphere of radio $R$. The Eq.
(\ref{eq:2cont}) is the accessible volume for a particle in a pore
with inclusion distance $PA$ (eventually $PA=P-A$), which is related
to the second coefficient of the pressure virial series \cite{HS_vir4Poly}
for a mixture of two HS gases. The CI of two spheres with a repulsion
distance $PA$ (eventually $PA=P+A$) is given by Eq. (\ref{eq:2dash}),
where $V_{\infty}$ is the volume of the total space. These obviously
are not novel results.

\subsection{Two spheres in a pore\label{sub:Two-spheres-in}}

Now we focus on the two spheres in a pore problem and then we need to
integrate the 2-in-pore graph (left-hand side graph on second row of
Eq. (\ref{eq:tri1})). Before integral evaluation we may state that
some set of values $\{ AB,PB,PA \}$ produce a zero sized graph, since
$A$ and $B$ spheres may not enter into the pore $P$. If $AB> PA+PB$,
then Eqs. (\ref{eq:ebond}) and (\ref{eq:fbond}) imply that
$r_{AB}> r_{PA}+r_{PB}$ thus avoiding the triangular relation and
therefore the 2-in-pore graph becomes zero. From now on we will
assume $AB\leq PB+PA$ and $PB\geq PA$, which is compatible with a
pore of radius $P$ that contains two particles of radius $A$ and $B$
($P\geq A+B$) with $A\geq B$.

For the calculus of density distribution it is useful introducing the
function $Z(r,R_{1},R_{2})$ \cite{HS_vir4Poly}, which represents
the volume of intersection of two spheres of radius $R_{1}$ and $R_{2}$
separated by a distance $r$ (assuming $R_{1}\geq R_{2}$)
\begin{equation}
\begin{array}{ccc}
Z(r,R_{1},R_{2}) & = & \left\{ \begin{array}{ll}
Sp(R_{2})\; & \mathrm{for}\:\, r\leq R_{1}-R_{2}\: \\
I_{2}(r,R_{1},R_{2})\; & \mathrm{other}\: \\
0\; & \mathrm{for}\:\, r>R_{1}+R_{2}\,,\end{array}\right.\end{array}
\label{eq:Zfunc}\end{equation}
\begin{equation}
I_{2}(r,R_{1},R_{2})=\frac{\pi}{12\, r}(R_{1}+R_{2}-r)^{2}(r^{2}-3(R_{1}-R_{2})^{2}+2r\,(R_{1}+R_{2}))\:.
\label{eq:I2func}\end{equation}
The function $I_{2}(r,R_{1},R_{2})$ is the volume in the partially
overlapping configuration, and is symmetric through the permutation
$R_{1}\leftrightarrow R_{2}$. Other functions related with $Z(r,R_{1},R_{2})$
are summarized in Table \ref{cap:Ztable} (the full extended form of
Table \ref{cap:Ztable} appears in Appendix \ref{Appendix-A}).
This set of functions is related with the integration of one spherical
body linked with other two. As we have two types of bonds $e$ and
$f$ there are four different Zeta functions. %
\begin{table}
\begin{center}\begin{tabular}{|c|c|c|c|c|}
\hline 
&
\raisebox{-18pt}{\psfig{figure=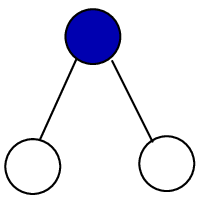,width=1.5cm}}&
\raisebox{-18pt}{\psfig{figure=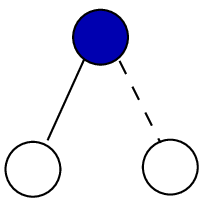,width=1.5cm}}&
\raisebox{-18pt}{\psfig{figure=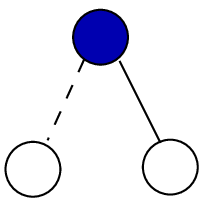,width=1.5cm}}&
\raisebox{-18pt}{\psfig{figure=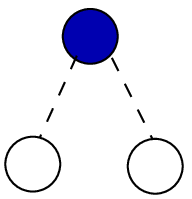,width=1.5cm}}\tabularnewline
&
$Z\equiv Z(r,R_{1},R_{2})$&
$Z_{\mathrm{I}}'(r,R_{1},R_{2})$&
$Z_{\mathrm{II}}'(r,R_{1},R_{2})$&
$Z''(r,R_{1},R_{2})$\tabularnewline
\hline
\hline 
HS&
&
$Sp(R_{1})-Z$&
$Sp(R_{2})-Z$&
$V_{\infty}-Sp(R_{1})-Sp(R_{2})+Z$\tabularnewline
\hline 
HD&
&
$Cr(R_{1})-Z$&
$Cr(R_{2})-Z$&
$V_{\infty}-Cr(R_{1})-Cr(R_{2})+Z$\tabularnewline
\hline
\end{tabular}\end{center}

\caption{Relation between the set of zeta functions \{$Z$, $Z_{\mathrm{I}}'$,$Z_{\mathrm{II}}'$,$Z''$\}
and the integration of one dark particle linked with two other white
particles. Note that in $Z_{\mathrm{I}}'$ and $Z_{\mathrm{II}}'$,
$R_{1}$ is the repulsion distance with the left white sphere, while
$R_{2}$ corresponds to the right one, with $R_{1}\geq R_{2}$.\label{cap:Ztable}}
\end{table}
 The use of these functions accounts us to write down two relevant
distribution functions in a simple way: the pair distribution function
$g_{2}(r_{AB})$ (in which the position of the pore center was integrated)
and the one body distribution function of one particle $\rho_{1}(r_{A})$.
Both, $r_{A}$ and $r_{B}$ have their origin in the center of the
pore
\begin{equation}
g_{2}(r_{AB})=\left(Q_{2}^{P}\right)^{-1}e_{AB}\: Z(r_{AB},PB,PA)\,,
\label{eq:g2}\end{equation}
\begin{equation}
\rho_{1}(r_{A})=\left(Q_{2}^{P}\right)^{-1}(-f_{PA})\,\times\left\{ \begin{array}{ll}
Z'_{\mathrm{I}}(r_{A},PB,AB)\; & \mathrm{for}\:\, AB\leq PB\, \\
Z'_{\mathrm{II}}(r_{A},AB,PB)\; & \mathrm{for}\:\, AB> PB\,,\end{array}\right.
\label{eq:rho1A}\end{equation}
\begin{equation}
\rho_{1}(r_{B})=\left(Q_{2}^{P}\right)^{-1}(-f_{PB})\,\times\left\{ \begin{array}{ll}
Z'_{\mathrm{I}}(r_{B},PA,AB)\; & \mathrm{for}\:\, AB\leq PA\, \\
Z'_{\mathrm{II}}(r_{B},AB,PA)\; & \mathrm{for}\:\, AB> PA\,,\end{array}\right.
\label{eq:rho1B}\end{equation}
where $\rho_{1}(r_{B})$ is equivalent to $\rho_{1}(r_{A})$ through
the transformation $A\leftrightarrow B$ in Eq. (\ref{eq:rho1A}).
Performing the complete integration (for example, in Eq. (\ref{eq:g2})),
we found the CI of two HS in a pore
\begin{equation}
Q_{2}^{P}=\int_{AB}^{PA+PB}4\pi\, r_{AB}^{2}\, Z(r_{AB},PB,PA)\, dr_{AB}\:,
\label{eq:QP2HS0}\end{equation}
\begin{equation}
Q_{2}^{P}=\left\{ \begin{array}{ll}
\frac{\pi^{2}}{18}\left(PA+PB-AB\right)^{3}\left(AB^{3}+3AB^{2}(PA+PB)\right.\\
\left.\:-(PA+PB+3AB)((PA-PB)^{2}-2PA\, PB)\right)\; & \mathrm{for}\:\, AB\geq PB-PA\, \\
\\\frac{16\pi^{2}}{9}PA^{3}\left(PB^{3}-AB^{3}\right)\; & \mathrm{for}\:\,
AB< PB-PA\,,\end{array}\right.
\label{eq:QP2}\end{equation}
which is valid as long as $AB\leq PA+PB$, whereas $Q_{2}^{P}=0$
if $AB>PA+PB$. There are explicit relations between the non additive,
the additive, and the additive-in-pore systems. These are expressed
in Eqs. (\ref{eq:mapadd}) and (\ref{eq:mapsus}) respectively, which
allow the map from one to other system when the triangular relation
is fulfilled
\begin{equation}
\begin{array}{lclclcl}
\, \, AB & = & A+B & & 2A & = & AB+PA-PB\\
\, \, PB & = & P+B \quad\; & & 2B & = & AB-PA+PB\\
\, \, PA & = & P+A & & 2P & = & -AB+PA+PB\,,
\end{array}
\label{eq:mapadd}\end{equation}
\begin{equation}
\begin{array}{lclclcl}
AB & = & A+B & & 2A & = & AB-PA+PB\\
PB & = & P-B &\quad & 2B & = & AB+PA-PB\\
PA & = & P-A & & 2P & = & AB+PA+PB\,.
\end{array}
\label{eq:mapsus}\end{equation}
Thus, using Eq. (\ref{eq:mapsus}) we may evaluate the $CI$ of two additive spheres in a pore
\begin{equation}
Q_{2}^{P}=\left(\frac{4\pi}{3}\right)^{2}\left(P-A-B\right)^{3}\left(P^{3}+3P\, A\,
B-A^{3}-B^{3}\right)\:,
\label{eq:QP2_addHS}\end{equation}
above expression is valid for $P\geq A+B$, whereas $Q_{2}^{P}=0$ if 
$P<A+B$. With the aim of check, using Eqs. (\ref{eq:tri1}, \ref{eq:QP2}, \ref{eq:mapadd})
we have obtained the third virial coefficient in the additive polydisperse
system \cite{HS_vir3Poly,HS_vir4Poly} the right-hand side graph on second
row of Eq. (\ref{eq:tri1}). In addition, in the case of equal sized spheres
with radii $A$ we obtain
\begin{equation}
Q_{2}^{P}=\left(\frac{4\pi}{3}\right)^{2}\left(P-2A\right)^{3}\left(P^{3}+3\: A^{2}P-2A^{3}\right)\:,
\label{eq:QP2_eqHS}\end{equation}
\begin{equation}
Q_{2}^{P}=\left(\frac{4\pi}{3}\right)^{2}\left(P-A\right)^{6}-\left(\frac{4\pi}{3}\right)^{2}(2A)^{3}\left(P-A\right)^{3}+\pi^{2}(2A)^{4}\left(P-A\right)^{2}-\frac{\,\pi^{2}}{18}(2A)^{6}\:.
\label{eq:QP2_eqHS2}\end{equation}
These two equations are equivalent. In Eq. (\ref{eq:QP2_eqHS}) it
seems clear the existence of a root on $Q_{2}^{P}$ at $P=2A$. On
the other hand, Eq. (\ref{eq:QP2_eqHS2}) shows the volume and area
dependence, being $V=(4\pi/3)(P-A)^{3}$ and $Ar=4\pi(P-A)^{2}$.
A simple relation connects $Q_{2}^{P}$ with the second cluster integral
$b_{2}(V)$ (which was stated in the second row of Eq. (\ref{eq:tri1}));
due to $b_{2}(V)$ has been analytically evaluated in previous works
\cite{HS_inhom} we were able to check the validity of our result.
Distribution functions for two additive spheres in a pore are shown
in Appendix \ref{Appendix-B}.

\subsection{Two disks in a pore}

The two dimensional problem of two HD in a circular pore is very similar
to the problem of two HS in a spherical pore in three dimensions.
Therefore we will briefly outline the aforementioned analysis concerning
HD. In the two particle system or one disk in a pore, we obtain
\begin{equation}
\raisebox{-1pt}{\psfig{figure=2fbis.eps,width=1.5cm}}=Cr(PA)\;,
\label{eq:2cont_D}\end{equation}
\begin{equation}
\raisebox{-1pt}{\psfig{figure=2ebis.eps,width=1.5cm}}=V_{\infty}-Cr(PA)\;,
\label{eq:2dash_D}\end{equation}
\begin{equation}
Cr(R)=\pi R^{2}\;,
\label{eq:CrAr}\end{equation}
where $Cr(R)$ is the surface area of the circle with radio $R$.
To maintain an unified point of view for both systems, HD and HS,
we name the measure of the total space $V_{\infty}$ (although it
is actually an area). The Eq. (\ref{eq:2cont_D}) is the accessible
area for a particle in a pore with an inclusion distance $PA$ (eventually
$PA=P-A$), which is related to the second coefficient of the pressure
virial series on a mixture \cite{HS_vir4Poly}. The configuration
integral of two disks with repulsion distance $PA$ (eventually $PA=P+A$)
is given by Eq. (\ref{eq:2dash_D}). For the two disk in a pore problem
we have
\begin{equation}
\begin{array}{ccc}
Z(r,R_{1},R_{2}) & = & \left\{ \begin{array}{ll}
Cr(R_{2})\; & \mathrm{for}\:\, r\leq R_{1}-R_{2}\: \\
I_{2}(r,R_{1},R_{2})\; & \mathrm{other}\: \\
0\; & \mathrm{for}\:\, r>R_{1}+R_{2}\:,\end{array}\right.\end{array}
\label{eq:ZfuncDisk}\end{equation}
\begin{equation}
\begin{array}{ccl}
I_{2}(r,R_{1},R_{2}) & = & R_{1}^{2}\, arccos[\frac{r^{2}+R_{1}^{2}-R_{2}^{2}}{2\, r\, R_{1}}]+R_{2}^{2}\, arccos[\frac{r^{2}-R_{1}^{2}+R_{2}^{2}}{2\, r\, R_{2}}]\\
\\ &  & -\frac{1}{2}\:\sqrt{(r+R_{1}+R_{2})(-r+R_{1}+R_{2})(r-R_{1}+R_{2})(r+R_{1}-R_{2})}\:.\end{array}
\label{eq:I2funcDisk}\end{equation}
The function $I_{2}(r,R_{1},R_{2})$ is the surface area in the partially
overlapping configuration. The set of functions related with $Z(r,R_{1},R_{2})$
for disks are summarized in Table \ref{cap:Ztable} (see the full
extended form in Appendix \ref{Appendix-A}).
On terms of the set of $Z$ functions it is possible to write down
the pair distribution function $g_{2}(r_{AB})$, and the one body
distribution function $\rho_{1}(r_{A})$
\begin{equation}
g_{2}(r_{AB})=\left(Q_{2}^{P}\right)^{-1}e_{AB}\: Z(r_{AB},PB,PA)\:,
\label{eq:g2_HD}\end{equation}
\begin{equation}
\rho_{1}(r_{A})=\left(Q_{2}^{P}\right)^{-1}(-f_{PA})\,\times\left\{ \begin{array}{ll}
Z'_{\mathrm{I}}(r_{A},PB,AB)\; & \mathrm{for}\:\, AB\leq PB\; \\
Z'_{\mathrm{II}}(r_{A},AB,PB)\; & \mathrm{for}\:\, AB> PB\;,\end{array}\right.
\label{eq:rho1A_HD}\end{equation}
where $\rho_{1}(r_{B})$ is equal to $\rho_{1}(r_{A})$ under the
transformation $A\leftrightarrow B$ in Eq. (\ref{eq:rho1A_HD}).
Both, $r_{A}$ and $r_{B}$, have their origin in the center of the
pore. Performing the complete integration of Eq. (\ref{eq:g2_HD}),
for example, we find the partition function for two disks in a pore
\begin{equation}
Q_{2}^{P}=\int_{AB}^{PA+PB}2\pi\, r_{AB}\, Z(r_{AB},PB,PA)\, dr_{AB}\;,
\label{eq:QP2HD0}\end{equation}
\begin{equation}
Q_{2}^{P}=\left\{ \begin{array}{lcl}
\;\pi\,\kappa\,\left(PB^{2}+PA^{2}+AB^{2}\right)\\
+\pi\, PB^{2}PA^{2}(\pi-arccos[\frac{PB^{2}+PA^{2}-AB^{2}}{2\, PB\, PA}])\\
-\pi\, PB^{2}AB^{2}arccos[\frac{PB^{2}-PA^{2}+AB^{2}}{2\, PB\, AB}]\\
-\pi\, PA^{2}AB^{2}arccos[\frac{-PB^{2}+PA^{2}+AB^{2}}{2\, PA\, AB}]\,
&  & \mathrm{for}\:\, AB\geq PB-PA\, \\
\\\;\pi^{2}PA^{2}\left(2PB^{2}-AB^{2}\right)\,
&  & \mathrm{for}\:\, AB<PB-PA\;,\end{array}\right.\:
\label{eq:QP2_HD}\end{equation}
\begin{equation}
\kappa=\frac{1}{4}\sqrt{\left(PB+PA+AB\right)\left(-PB+PA+AB\right)\left(PB-PA+AB\right)\left(PB+PA-AB\right)}\;,
\label{eq:kTrAr}\end{equation}
where $\kappa$ represents the area of the triangle with sides $PB$,
$PA$, $AB$. The arguments of the arccosine function in Eq. (\ref{eq:QP2_HD})
are the internal angles of this triangle and their sum is $\pi$.
In particular, for the additive in pore system with different sized
disks, the previous equation can be simplified
\begin{equation}
\begin{array}{ccl}
Q_{2}^{P} & = & 2\pi\kappa\left(P^{2}+A^{2}+B^{2}-P\,(A+B)+A\, B\right)\\
\\ & & +\pi(P-B)^{2}(P-A)^{2}(\pi-arccos[\frac{P^{2}-P\,(A+B)-A\, B}{(P-B)(P-A)}])\\
\\ &  & -\pi(P-B)^{2}(A+B)^{2}arccos[\frac{P\,(A-B)+B^{2}+A\, B}{(P-B)(A+B)}]\\
\\ & & -\pi(P-A)^{2}(A+B)^{2}arccos[\frac{-P\,(A-B)+A^{2}+A\, B}{(P-A)(A+B)}]\:,\end{array}
\label{eq:QP2_addHD}\end{equation}
\begin{equation}
\kappa=\sqrt{P\, B\, A\left(P-A-B\right)}\;.
\label{eq:kTrAr_add}\end{equation}
The Eq. (\ref{eq:QP2_addHD}) applies in the case $P\geq A+B$, whereas
$Q_{2}^{P}=0$ if $P<A+B$. Using Eqs. (\ref{eq:tri1}, \ref{eq:mapadd},
and \ref{eq:QP2_addHD}) the third virial coefficient is correctly
obtained in the additive polydisperse system \cite{HD_vir3Poly}.
By evaluating the above expressions for two disks with equal radii,
we find
\begin{equation}
\begin{array}{ccl}
Q_{2}^{P} & = & 2\pi\, A\:\sqrt{P\,\left(P-2A\right)}\left((P-A)^{2}+2A^{2}\right)\\
\\ &  & +\,2\pi\,(P-A)^{2}\left((P-A)^{2}-(2A)^{2}\right)arccos[\frac{A}{P-A}]\:,\end{array}
\label{eq:QP2_eqHD}\end{equation}
or
\begin{equation}\begin{array}{ccl}
Q_{2}^{P} & = & \left(\frac{128}{15}\,\sqrt{2}\,\pi\,
A^{3/2}\right)\left(P-2A\right)^{5/2} \\
\\ & & \times\,\left(1+\frac{19}{28\, A}\left(P-2A\right)+\frac{9}{224\,A^{2}}\left(P-2A\right)^{2}+
O_{3}\left(P-2A\right)\right)\:,\end{array}
\label{eq:QP2_eqHD2}\end{equation}
Eq. (\ref{eq:QP2_eqHD}) has a root at $P=2A$. Thus, in Eq. (\ref{eq:QP2_eqHD2})
we study the local behaviour of Eq. (\ref{eq:QP2_eqHD}) near its
root. Distribution functions for two additive disks
in a pore is discused in Appendix \ref{Appendix-B}.

\section{Results\label{sec:Results}}

Let us make a few remarks on the new results obtained in this work. The
configuration integral in Eqs. (\ref{eq:QP2}, \ref{eq:QP2_HD}) are
analytical functions of the system parameters in the domain ${AB,PA,PB}>0$ 
except when we focus on the non additive regime for $AB=PB-PA$. From here
on we will restrict ourselves to $AB\leq PA+PB$. For HS the configuration
integral is polynomial, although it is not true for HD. In the two
additive bodies in a pore system, the CI Eqs. (\ref{eq:QP2_addHS},
\ref{eq:QP2_addHD}) become null at close packing pore size $P_{CP}=A+B$,
even though systems do not lose the rigid rotation degrees of freedom.
Over all density range $P\in(P_{CP}\,,+\infty)$ CIs are monotonic analytic
functions, therefore, none ergodic/non-ergodic transition happens and
no van der Waals loop is expected \cite{HD_2inBox,HD_3inBox}.

As we are interested in thermodynamic properties we analyze the Free
energy of the system. Taken into account the kinetic factor, which
contains the temperature dependence and usual thermodynamic relations
we obtain the entropy $S$
\begin{equation}
S/k=-\beta\, F+\beta\, U=ln(Q_{2}^{P})\;,
\label{eq:FreeEn}\end{equation}
where $U$ is the energy of the ideal gas ($\beta U=2,\,3$ for two
HD and HS, respectively). An important result derived from Eqs. (\ref{eq:QP2_addHS},
\ref{eq:QP2_addHD}) is that the Free Energy diverges logarithmically
at $P_{CP}$. In order to obtain the pressure we need to define the
system volume $V_{sys}$, both magnitudes are related by
\begin{equation}
\beta\, P_{W}=\left.\frac{d\, ln(Q_{2}^{P})}{d\,
V_{sys}}\right|_{\beta}=\left.\frac{d\, ln(Q_{2}^{P})}{d\, P}
(V'_{sys})^{-1}\right|_{\beta}\;,
\label{eq:Pressure0}\end{equation}
where $V'_{sys}$ is the area of the surface enclosing $V_{sys}$. A
particular volume definition implicitly induce a surface definition and
then only modifies the surface area in Eq. (\ref{eq:Pressure0}). Although
we are considering a small and very simple system composed by solely two
hard spherical particles in a pore, it is not obvious which is the system
volume. In principle, three different volumes may be considered, the volume
accessible to the particle $A$, that is $V_{A}$; the same for the $B$
particle i.e. $V_{B}$ (see Eqs. (\ref{eq:2cont}, \ref{eq:2cont_D})); and
the volume of the free space in the pore. The later may be interpreted
as the accessible volume for a particle with a vanishing radio and
does not depend on the size of the chosen particle. We propose that
$V_{A}$ and $V_{B}$ are the relevant volumes. In such a case the
pressure on the wall became
\begin{equation}
\beta\, P_{W}=\left.\frac{d\, ln(Q_{2}^{P})}{d\, V_{A}}\right|_{\beta , V_{B}}+\left.\frac{d\, ln(Q_{2}^{P})}{d\, V_{B}}\right|_{\beta
,V_{A}}\;,
\label{eq:Pressure}\end{equation}
\begin{equation}
\beta\, P_{W}=\rho_{1}(r_{A}=P-A)+\rho_{1}(r_{B}=P-B)\;,
\label{eq:PressureRho}\end{equation}
where Eqs. (\ref{eq:Pressure}, \ref{eq:PressureRho}) are generalizations
of Dalton's Law, and the equivalence between them is trivial.

For the sake of clarity from here on we will analyze the system of
two equal sized particles, i.e. $B=A$ and $P_{CP}=2A$. An interesting
point about Eq. (\ref{eq:QP2_eqHS2}) for two HS, is the fact that
it has the simplest possible shape. Indeed, each term has a very simple
mean. The first term refers to the ideal gas, while the second term
is clearly related to the first virial series correction (due to the
interaction between particles) both of them appear in the homogeneous
two body system. The third term is the product between the area of
the system and the second surface virial coefficient at a planar hard
wall $W_{2}$ \cite{Bell,SteckiSoko}. Finally the fourth term forces
the root of CI at $P=P_{CP}$. From Eqs. (\ref{eq:QP2_eqHS}) and
(\ref{eq:QP2_eqHD2}) the CI goes to zero at $P_{CP}$ with multiplicity
$3$ and $5/2$ for HS and HD, respectively, and its value for two
HD in a square box is $4$ \cite{HD_2inBox}. The mean of such values
are probably related to the number of degrees of freedom lost at the
\textit{ultimate solid} or $cp$ density. The logarithmic divergence
of the Free Energy and Eq. (\ref{eq:Pressure}) induce an order one
pole at the maximum density. The equations of state for HS and HD
are, respectively,
\begin{equation}
\beta P_{W}=\frac{1}{4\pi(P-1)^{2}}\left(\frac{3}{(P-2)}+\frac{5}{4}-\frac{9}{16}(P-2)+O_{2}(P-2)\right)\;,
\label{eq:PrHS}\end{equation}
and
\begin{equation}
\beta P_{W}=\frac{1}{2\pi(P-1)}\left(\frac{5/2}{(P-2)}+\frac{19}{28}-\frac{149}{392}(P-2)+O_{2}(P-2)\right)\;.
\label{eq:PrHD}\end{equation}
Here, we choose $A$ as the unit length. Note that Eqs. (\ref{eq:PrHS})
and (\ref{eq:PrHD}) are truncated forms of the equation of state
but the exact complete expressions can be easily derived from Eqs.
(\ref{eq:QP2_eqHS}, \ref{eq:QP2_eqHD}, \ref{eq:Pressure}). It seems
that some of the afore analyzed properties about the \textit{ultimate
solid} are valid for any finite number of particles $N$. That is,
the CI must go to zero at some $P=P_{CP}(N)$ value with a multiplicity
$\alpha_{-1}(N)$, which is a positive real number. This characteristic
implies the logarithmic divergence of the Free Energy and induces an
order one pole in $P_{W}$ at the maximum density. Then we conjecture
that the equation of state near the \textit{ultimate solid} density,
takes the form

\begin{equation}
\beta P_{W}=\frac{1}{\Omega_{D}\,(P-1)^{D-1}}\left(\frac{\alpha_{-1}}{(P-P_{CP})}+\alpha_{0}+\alpha_{1}(P-P_{CP})+O_{2}(P-P_{CP})\right)\;,
\label{eq:PrN}\end{equation}
where $D$ is the dimensionality of the system and $\Omega_{D}$ is
the solid angle integral. This last equation can be rewritten in the
following way
\begin{equation}
\begin{array}{ccl}
\beta P_{W}\frac{V}{N} & = & \frac{\alpha_{-1}}{N\, D}\,\left(1-(V_{o}/V)^{1/D}\right)^{-1}+
\frac{\alpha_{0}(P_{CP}-1)}{N\, D} \\
\\ & &+\frac{\alpha_{0}+\alpha_{1}(P_{CP}-1)}{N\, D}(P-P_{CP})+O_{2}(P-P_{CP})\:,\end{array}
\label{eq:PrN2}\end{equation}
where $V=\Omega_{D}\,(P-1)^{D}/D$ and $V_{o}=\Omega_{D}\,(P_{CP}-1)^{D}/D$.
It is interesting to mention that first term of the Eq. (\ref{eq:PrN2})
is equal to the equation of state for the bulk solid phase of HS and
HD systems at the highest density limit \cite{HSsolid}.

\begin{figure}[hd]
\begin{center}\includegraphics[%
  clip,
  width=7cm,
  keepaspectratio]{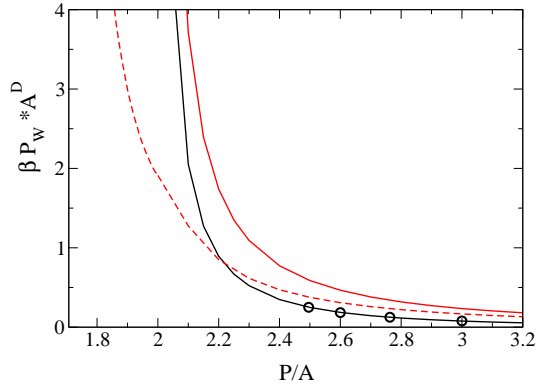}\end{center}
\caption{(Color online) Equation of state for two HD and HS. Continuous curves correspond
to HD (up) and HS (down) calculated in the present work. For comparison,
we also include the results for: two HD in a square box \cite{HD_2inBox}
(dashed line), and Monte Carlo simulation for two HS in a spherical
pore \cite{HSp_2inSphMC} (circles).\label{cap:EOS}}
\end{figure}

As was mentioned before, CIs are monotonic analytic functions, then
the pressure is a monotonic analytic function of $P\in(2,+\infty)$
and does not develop a van der Waals loop. This is not the case for
two and three HD in a rectangular box \cite{HD_2inBoxMD,HD_2inBox,HD_3inBox}.
In Fig. \ref{cap:EOS} we display the equations of state (\ref{eq:PrHS},
\ref{eq:PrHD}) as a function of pore size, both of them diverge at $P=2$.
For comparison, the equation of state for two HD system confined
in a square box pore is also plotted, there the half side of the square
box pore was taken as the size parameter $P$. The equation of state
for two HD in a square box has two branches that are joined on a non
analytic point at $P=2$, while the divergence appears at $P=P_{CP}=1+1/\sqrt{2}$
and takes place when the discs are fitted on the two opposite vertices
of the box. These two equations of state for two HD on different shaped
pores converge as the pore size grows. Present results of $\beta P_{W}$
for HS perfectly agree with the Monte Carlo results \cite{HSp_2inSphMC}
(extracted from Fig. 1 on that work) which were obtained by numerical
integration of the same equations.

The density distribution of one HS particle corresponding to a system with 
two equal sized HS (see Eq. (\ref{eq:rho1A_eqsust}) in the Appendix
\ref{Appendix-B}) has been plotted on Fig. \ref{cap:rho}. Density
profiles for several radii of the pore $P$ are shown in this figure by a
solid line, each curve has a maximum at $r=P-1$ and fall off to zero
discontinuously for $r>P-1$. Density distributions change gradually from
the complete radial localization, when $P=P_{CP}=2$ to the quasi uniform
distribution at $P=4$. For a pore size $P>3$ the density at $r=0$ becomes
non null and a constant density plateau develops for $0<r<P-3$. The dashed
and dot-dashed curves show the evolution of the minimum (at $r=0$)
and the maximum (at $r=P-1$) values of the density profiles as a
function of pore radio $P$. The dotted curve indicates once again
the contact density $\rho(P,r=P-1)$, which is related to the pressure
of the system by Eq. (\ref{eq:PressureRho}), but now as a function
of $P-1$. These three lines are functions of pore size, and refer
to the top abscissa axis. We find that all these properties are also
exhibited by the two HD in a circular pore system. We can mention
that at $P=3.4069$ the central density reaches its maximum value,
which is close but greater than half of the density at contact. 
\begin{figure}[hd]
\begin{center}\includegraphics[%
  clip,
  width=7cm,
  keepaspectratio]{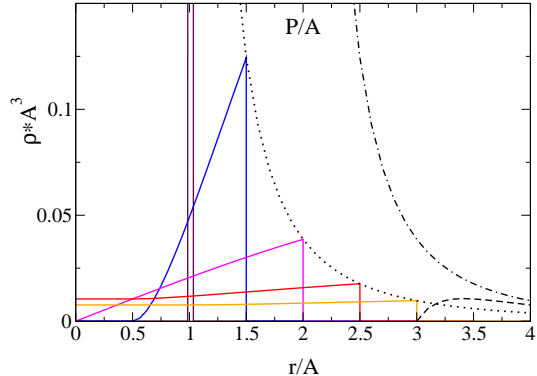}\end{center}
\caption{(Color online) Density profiles for two HS system as a function of
\textit{r} for various \textit{P} values. Both, \textit{P} and \textit{r}
are in units of \textit{A}. Continuous lines indicate results for pore
radii $P=2.01,\:2.5,\:3,\:3.5,\:4$. Other curves are referred to top
abscissa axis. Dashed curve shows density at central point $\rho(P,r=0)$,
whereas dot-dashed curve corresponds to density at contact $\rho(P,r=P-1)$.
The dotted curve is the shifted dot-dashed one (see text).\label{cap:rho}}
\end{figure}

The complete dependence of the second cluster integral $b_{2}(V)$
for a system of particles contained into a spherical vessel is related
to the second virial coefficient by $B(V)=-b_{2}(V)$. From Eq. (\ref{eq:tri1})
we obtain\begin{equation}
V\, b_{2}(V)=\frac{1}{2}\left(Q_{2}^{P}-V^{2}\right).\label{eq:b2deQ}\end{equation}
Using Eq. (\ref{eq:QP2_eqHS}) we derived $b_{2}(V)$ for two equal
HS in a spherical pore,

\begin{equation}
V\, b_{2}(V)=b_{2}V+2\pi\, A^{4}Ar-(\frac{4}{3}\pi)^{2}A^{6}\:,
\label{eq:b2Sph}\end{equation}
where $V=\left(\frac{4\pi}{3}\right)\left(P-A\right)^{3}$, $b_{2}=-\frac{2\pi}{3}(2A)^{3}$,
and $Ar=4\pi\left(P-A\right)^{2}$. As was discussed before (in Sec.
\ref{sub:Two-spheres-in}) this expression was previously calculated
in literature \cite{HS_inhom}. For the first time, we present the
complete dependence of $b_{2}(V)$ for a HD system into a circular
vessel. Then, using Eqs. (\ref{eq:QP2_eqHD}) and (\ref{eq:b2deQ})
\begin{equation}
V\, b_{2}(V)=b_{2}V+\frac{1}{3}(2A)^{3}Ar-\frac{1}{60}\,\pi(2A)^{5}\frac{1}{\left(P-A\right)}+O_{3}\left(\frac{1}{P-A}\right)\:,
\label{eq:b2Disc}\end{equation}
where $V=\pi\left(P-A\right)^{2}$, $b_{2}=-\frac{1}{2}\pi(2A)^{2}$,
and $Ar=2\pi\left(P-A\right)$. Actually $V$ is the accessible area
(see Eq. (\ref{eq:CrAr})) and $Ar$ is its perimeter. Although in
Eq. (\ref{eq:b2Disc}) we choose express $b_{2}(V)$ as a truncated
series expansion of the complete expression, the exact form of $b_{2}(V)$
(for two HD and HS with equal or unequal size) may be obtained with
the same procedure. The factor multiplying $Ar$ is the second virial
coefficient of the surface tension (in fact the linear tension) $W_{2}$
\cite{Bell,SteckiSoko}, as far as we know it was not previously evaluated.
The third term is the first finite size-curvature correction and it
has a minus one power on $\left(P-A\right)$, which is two units smaller
than the former term as happens in HS at Eq. (\ref{eq:b2Sph}). All
the forthcoming terms of the series (not shown in Eq. (\ref{eq:b2Disc}))
are negatives with odd degree in $\left(P-A\right)^{-1}$. The expansion
of Eq. (\ref{eq:b2Disc}) converges so quickly that, if we truncate
at fifth order, the deviation is smaller than one percent in the worst
case for $P=2A$, when the system attains its highest density.

We can mention that CI of two Hard Rod system confined into a segment
(the one dimensional equivalent of two HD and HS in a spherical pore)
is easy to evaluate analytically \cite{HR,HRbis}. Then, this work
provides a set of analytic properties for dimensions $D=1,\,2$ and
$3$, and any general approximate theory for inhomogeneous confined
liquid phase \cite{Taraz} may be compared with these exact results.

\section{Conclusions\label{sec:Conclusions}}

We have established a deep relation between CI of a N-polydisperse
system of hard spheres, CI of a (N-1)-polydisperse system in a spherical
shaped pore, and Mayer type diagrams \cite{BookHansen,BookHill}.
The relation valid in any dimension was used in the present work only
for checking purposes. A similar relation was stated previously in
the Gran Canonical Ensemble \cite{HS_inhom}.

The canonical partition and distribution functions of two HS and two
HD in a spherical shaped pore were analyzed in an exact analytic framework.
The analysis presented here was made assuming a very general system
of two HD and HS, with different sizes and non additive potential.
The obtained exact CI expression allow us to evaluate the thermodynamic
observables of the system (Free Energy, Energy, Entropy, Pressure,
etc) \cite{BookHansen,BookHill}. In addition, we investigate the
analytic properties of CI and the equation of state at the \textit{ultimate
solid} or $cp$ density, and we compare our results with the known
exact result of two HD confined in a square box. Based on this study
we present a general proposal for the equation of state at the \textit{ultimate
solid}, in a system of N hard spherical particles contained into a
pore. We also find the second virial coefficient for a HD system in
a circular pore. Future works will involve the analytical evaluation
of the CI for two particles in other simple shaped pores, and the
analytic study of the CI for three particles in spherical pore.

\section*{Acknowledgements}

I am grateful to Dr. Leszek Szybisz and Dr. Gabriela Castelletti for
all their suggestions. This work was supported in part by the Ministry
of Culture and Education of Argentina through Grants CONICET PIP No.
5138/05 and UBACyT No. X298.

\appendix

\section{Appendix A: Expanded definition of $Z'_{\mathrm{I}}$, $Z'_{\mathrm{II}}$
and $Z''$\label{Appendix-A}}

In this appendix it is showed the expanded form of the $Z(r,R_{1},R_{2})$
related set of functions for spherical bodies. We name $D$ to the
dimensionality of space and we assume that $R_{1}\geq R_{2}$. In
the present notation an unified point of view for spheres and disks
is chosen. The family of $Z$ functions in arbitrary dimensions reads,

\begin{equation}
\begin{array}{ccc}
Z(r,R_{1},R_{2}) & = & \left\{ \begin{array}{ll}
I_{1}^{(D)}(R_{2})\; & \mathrm{for}\:\, r\le R_{1}-R_{2}\, \\
I_{2}^{(D)}(r,R_{1},R_{2})\; & \mathrm{other}\\
0\; & \mathrm{for}\:\, r>R_{1}+R_{2}\,, \end{array}\right.\end{array}
\label{eq:Z_D}\end{equation}
\begin{equation}
\begin{array}{ccc}
Z'_{\mathrm{I}}(r,R_{1},R_{2}) & = & \left\{ \begin{array}{ll}
I_{1}^{(D)}(R_{1})-I_{1}^{(D)}(R_{2})\; & \mathrm{for}\:\, r\le R_{1}-R_{2}\, \\
I_{1}^{(D)}(R_{1})-I_{2}^{(D)}(r,R_{1},R_{2})\; & \mathrm{other}\, \\
I_{1}^{(D)}(R_{1})\; & \mathrm{for}\:\, r>R_{1}+R_{2}\,,\end{array}\right.\end{array}
\label{eq:Zpi_D}\end{equation}
\begin{equation}
\begin{array}{ccc}
Z'_{\mathrm{II}}(r,R_{1},R_{2}) & = & \left\{ \begin{array}{ll}
0\; & \mathrm{for}\:\, r\le R_{1}-R_{2}\, \\
I_{1}^{(D)}(R_{2})-I_{2}^{(D)}(r,R_{1},R_{2})\; & \mathrm{other}\, \\
I_{1}^{(D)}(R_{2})\; & \mathrm{for}\:\, r>R_{1}+R_{2}\,,\end{array}\right.\end{array}
\label{eq:Zpii_D}\end{equation}
\begin{equation}
\begin{array}{ccc}
Z''(r,R_{1},R_{2}) & = & \left\{ \begin{array}{ll}
V_{\infty}-I_{1}^{(D)}(R_{1})\; & \mathrm{for}\:\, r\le R_{1}-R_{2}\, \\
V_{\infty}-I_{1}^{(D)}(R_{1})-I_{1}^{(D)}(R_{2})+I_{2}^{(D)}(r,R_{1},R_{2})\; & 
\mathrm{other}\, \\
V_{\infty}-I_{1}^{(D)}(R_{2})-I_{1}^{(D)}(R_{2})\; & \mathrm{for}\:\,
r>R_{1}+R_{2}\,,\end{array}\right.\end{array}
\label{eq:Zpp_D}\end{equation}
where $I_{1}^{(D)}(R_{1})$ is the volume of the D-sphere with radio
$R_{1}$ (the expression for $D=3$ and $D=2$ are in Eqs. (\ref{eq:SpVol},\ref{eq:CrAr})),
and $I_{2}^{(D)}(r,R_{1},R_{2})$ is the volume of intersection of
two D-spheres with radii $R_{1}$ and $R_{2}$, separated by a distance
$r$ in the partial overlapping configuration (the expression for
$D=3$ and $D=2$ are in Eqs. (\ref{eq:I2func},\ref{eq:I2funcDisk})).

\section{Appendix B: Distribution functions in the
additive-in-pore system\label{Appendix-B}}

Assuming an additive-in-pore system, i.e. $AB=A+B$, $PA=P-A$ and $PB=P-B$,
the distribution functions for two HD or HS into a spherical pore (from Eqs.
(\ref{eq:g2}, \ref{eq:rho1A}) and (\ref{eq:g2_HD}, \ref{eq:rho1A_HD})) take
the form
\begin{equation}
g_{2}(r_{AB})=\left(Q_{2}^{P}\right)^{-1}\times\left\{ \begin{array}{ll}
I^{(D)}_{2}(r_{AB},P-B,P-A)\; & \mathrm{if}\:\, A+B\leq r_{AB}\leq2P-(A+B)\, \\
0\; & \mathrm{if}\:\, r_{AB}<A+B\:\,\mathrm{or}\:\, r_{AB}>2P-(A+B)\,,\end{array}\right.
\label{eq:g2_addsust}\end{equation}
\begin{equation}
\rho_{1}(r_{A})=\left(Q_{2}^{P}\right)^{-1}\times\left\{ \begin{array}{ll}
max[I^{(D)}_{1}(P-B)-I^{(D)}_{1}(A+B),0]\;
& \mathrm{if}\:\,0\leq r_{A}\leq abs[P-A-2B]\, \\
I^{(D)}_{1}(P-B)-I^{(D)}_{2}(r_{A},P-B,A+B)\; &
\mathrm{if}\:\, abs[P-A-2B]< r_{A}\leq P-A\, \\
0\; & \mathrm{if}\:\, r_{A}>P-A\,,\end{array}\right.
\label{eq:rho1A_addsust}\end{equation}
the $\rho_{1}(r_{B})$ function can be found by the permutation of
symbols $A\leftrightarrow B$ in Eq (\ref{eq:rho1A_addsust}). If we have equal
radii spherical particles, expressions become
\begin{equation}
g_{2}(r_{AA})=\left(Q_{2}^{P}\right)^{-1}\times\left\{ \begin{array}{ll}
I^{(D)}_{2}(r_{AA},P-A,P-A)\; & \mathrm{if}\:\,2A\leq r_{AA}\leq2(P-A)\, \\0\;
& \mathrm{if}\:\, r_{AA}<2A\:\,\mathrm{or}\,\:
r_{AA}>2(P-A)\,,\end{array}\right.
\label{eq:g2_eqsust}\end{equation}
\begin{equation}
\rho_{1}(r_{A})=\left(Q_{2}^{P}\right)^{-1}\times\left\{ \begin{array}{ll}
max[I^{(D)}_{1}(P-A)-I^{(D)}_{1}(2A),0]\;
& \mathrm{if}\:\,0\leq r_{A}\leq abs[P-3A]\, \\
I^{(D)}_{1}(P-A)-I^{(D)}_{2}(r_{A},P-A,2A)\; & \mathrm{if}\:\, abs[P-3A]< r_{A}\leq P-A\, \\
0\; & \mathrm{if}\:\, r_{A}>P-A\,.\end{array}\right.
\label{eq:rho1A_eqsust}\end{equation}
 Although the existence of the exact one body distribution function
for two HS in a spherical pore is mentioned in \cite{Gonzal}, neither
explicit expression was shown, nor further analysis was performed
there.

\end{document}